\documentclass[10.75pt,preprint]{article}
\usepackage[english]{babel}
\usepackage{amsmath,amssymb,graphicx,calc,capt-of,ifthen,theorem,pifont,amsfonts}

\begin{document}

\title{$\tilde{\delta}$ Gravity,  $\tilde{\delta}$ matter and the accelerated expansion of the Universe}
\author{Jorge Alfaro\footnote{Facultad de F\'isica, Pontificia Universidad Cat\'olica de Chile. Casilla 306, Santiago, Chile. jalfaro@uc.cl.}   and Pablo Gonz\'alez\footnote{Departamento de F\'isica, FCFM, Universidad de Chile. Blanco Encalada 2008, Santiago, Chile. pgonzalez@ing.uchile.cl (Current Affiliation).} \footnote{Facultad de F\'isica, Pontificia Universidad Cat\'olica de Chile. Casilla 306, Santiago, Chile. pegonza2@uc.cl.} }
\maketitle

\section*{ABSTRACT.}

A gravitational field model based on two symmetric tensors, $g_{\mu \nu}$ and $\tilde{g}_{\mu \nu}$, is presented. In this model, new matter fields are added to the original matter fields, motivated by an additional symmetry ($\tilde{\delta}$ symmetry). We call them $\tilde{\delta}$ matter fields. We find that massive particles do not follow geodesics, while trajectories of massless particles are null geodesics of an effective metric. Then we study  the Cosmological case, where we get an accelerated expansion of the Universe without dark energy.\\

\section*{Introduction.}

Recent discoveries in cosmology have revealed that most part of matter is in the form of unknown matter, dark matter {\cite{DM New 1}}-{\cite{DM DE 1}}, and that the dynamics of the expansion of the Universe is governed by a mysterious component that accelerates its expansion, the so called dark energy {\cite{DM DE 2}}-{\cite{DM DE 4}}. That is the Dark Sector. Although GR is able to accommodate the Dark Sector, its interpretation in terms of fundamental theories of elementary particles is problematic {\cite{DM DE 5}}. Although some candidates exist that could play the role of dark matter, none have been detected yet. Also, an alternative explanation based on the modification of the dynamics for small accelerations cannot be ruled out {\cite{DM DE 6,DM DE 7}}. Dark energy can be explained if a small cosmological constant ($\Lambda$) is present. At early times, this constant is irrelevant, but at the later stages of the evolution of the Universe $\Lambda$ will dominate the expansion, explaining the observed acceleration. However $\Lambda$ is too small to be generated in quantum field theory (QFT) models, because $\Lambda$ is the vacuum energy, which is usually predicted to be very large {\cite{lambda problem}}.\\

One of the most important mysteries in cosmology and cosmic structure formation is to understand the nature of dark energy in the context of a fundamental physical theory {\cite{DE 1,DE 2}}. In recent years there has been various proposals to explain the observed acceleration of the Universe. They include some additional fields in approaches like quintessence, chameleon, vector dark energy or massive gravity; The addition of higher order terms in the Einstein-Hilbert action, like $f (R)$ theories and Gauss-Bonnet terms and finally the introduction of extra dimensions for a modification of gravity on large scales (See {\cite{DE 3}}).\\

Recently, in \cite{delta gravity}, a model of gravitation that is very similar to GR is presented, but works different at the quantum level. In that paper, we considered two different points. The first is that GR is finite on shell at one loop in vacuum {\cite{tHooft}}, so renormalization is not necessary at this level. The second is the $\tilde{\delta}$ gauge theories (DGT) originally presented in {\cite{Alfaro 2,Alfaro 3}}, where the main properties are: (a) A new kind of field $\tilde{\phi}_{I}$ is introduced, different from the original set $\phi_{I}$. (b) The classical equations of motion of $\phi_{I}$ are satisfied even in the full quantum theory. (c) The model lives at one loop. (d) The action is obtained through the extension of the original gauge symmetry of the model, introducing an extra symmetry that we call $\tilde{\delta}$ symmetry, since it is formally obtained as the variation of the original symmetry. When we apply this prescription to GR, we obtain $\tilde{\delta}$ Gravity. For these reasons, the original motivation was to develop the quantum properties of this model (See \cite{delta gravity}).\\

Before continuing we want to introduce  a word of caution. In what follows we want to study $\delta$-gravity as a  {\it classical effective} model and use it in Cosmology. This means to approach the problem  from the phenomenological side
instead of neglecting it {\it a priori} because it does not satisfy yet all the
properties of a fundamental quantum theory. Examples of this approach in the current literature are plentiful\cite{phantom 1,phantom 2,phantom 3,phantom 4,phantom 5,phantom 6}.

The nature of Dark Energy is such an important and difficult cosmological problem that cosmologists do not  expect to find a fundamental solution of it in one stroke and are open to explore new possibilities.

In \cite{DG DE,Paper DE} we presented truncated versions of $\tilde{\delta}$ Gravity applied to Cosmology. The $\tilde{\delta}$ symmetry was fixed in different ways in order to simplify the analysis of the model. The results were quite reasonable taking into account the simplifications involved. In this work we present the full fletged $\tilde{\delta}$ Gravity. In order to preserve the  $\tilde{\delta}$ symmetry we must introduce $\tilde{\delta}$ matter.\\

We will see that the main properties of this model at the classical level are: (a) We agree with GR, far from the sources. In particular, the causal structure of $\tilde{\delta}$ Gravity in vacuum is the same as in general relativity. (b) The necessary quantity of dark matter could be considerably less that we expected. (c) When we study the evolution of the Universe, it predicts accelerated expansion  without a cosmological constant or additional scalar fields. The Universe ends in a Big-Rip, similar to the scenario considered in \cite{phantom 1}-\cite{phantom 5}. (d) The scale factor agrees with the standard cosmology at early times and shows acceleration only at late times. Therefore we expect that primordial density perturbations should not have large corrections.\\

Moreover the value of the cosmological parameters improve greatly by the inclusion of  $\tilde{\delta}$ matter. For instance, the age of the Universe is much closer to the Planck satellite value now than the value we got in \cite{Paper DE}.\\

In \textbf{Section \ref{Sec: delta Gravity}}, we will present the $\tilde{\delta}$ Gravity action that is invariant under extended general coordinate transformation. We will find the equations of motion of this action. We will see that the Einstein's equations continue to be valid and we will obtain a new equation for $\tilde{g}_{\mu \nu}$. In these equations, two energy momentum tensors, $T_{\mu \nu}$ and $\tilde{T}_{\mu \nu}$, are defined. Additionally, we will derive the equation of motion for the test particle. We distinguish the massive case, where the equation is not a geodesic, and the massless case, where we have a null geodesic with an effective metric. In \textbf{Section \ref{Sec: Cosmological Case}}, we will study the cosmological case. We obtain the accelerated expansion of the universe assuming a universe without dark energy, i.e. only having non-relativistic matter and radiation which satisfy a fluid-like equation $p= \omega \rho$. We will see that the most relevant element is the fraction between radiation and non-relativistic matter density in the present, producing a Big-Rip. A preliminary computation was done in {\cite{DG DE}}, where an approximation is discussed. Later, in {\cite{Paper DE}}, we developed an exact solution of the equations consistent with the above assumptions. However, in both cases we assumed that we do not have $\tilde{\delta}$ matter, such that the new symmetry is broken.  In all these cases, the solution is used to fit the supernovae data and we obtained a physical reason for the accelerated expansion of the universe within the model: the existence of massless particles. If massless particles were absent, the expansion of the Universe would be the same as in GR without a cosmological constant.\\

It should be remarked that $\tilde{\delta}$ Gravity is not a metric model of gravity because massive particles do not move on geodesics. Only massless particles move on null geodesics of a linear combination of both tensor fields. Additionally, it is important to notice that we will work with the $\tilde{\delta}$ modification for General Relativity, based on the Einstein-Hilbert theory. From now on, we will refer to this model as $\tilde{\delta}$ Gravity.\\

\newpage

\section{\label{Sec: delta Gravity}$\tilde{\delta}$ Gravity.}

Using the prescription given by \textbf{Appendix A}, we will present the action of $\tilde{\delta}$ Gravity and then we will derive the equations of motion. Additionally, we will study the test particle action separately for massive and massless particles.

\subsection{\label{SubSec: Equations of Motion}Equations of Motion:}

Now, we are ready to study the modifications of gravity. For this, let us consider the Einstein-Hilbert Action:

\begin{eqnarray}
\label{EH action}
S_0 = \int d^4x \sqrt{-g} \left(\frac{R}{2\kappa} + L_M\right),
\end{eqnarray}

where $L_M = L_M(\phi_I,\partial_{\mu}\phi_I)$ is the lagrangian of the matter fields $\phi_I$. Using (\ref{Action}), this action becomes:

\begin{eqnarray}
\label{grav action}
S = \int d^4x \sqrt{-g} \left(\frac{R}{2\kappa} + L_M - \frac{1}{2\kappa}\left(G^{\alpha \beta} - \kappa T^{\alpha \beta}\right)\tilde{g}_{\alpha \beta} + \tilde{L}_M\right),
\end{eqnarray}

where $\kappa = \frac{8 \pi G}{c^2}$, $\tilde{g}_{\mu \nu} = \tilde{\delta}g_{\mu \nu}$ and:

\begin{eqnarray}
\label{EM Tensor}
T^{\mu \nu} = \frac{2}{\sqrt{-g}} \frac{\delta}{\delta g_{\mu \nu}}\left[\sqrt{-g} L_M\right] \\
\label{tilde L matter}
\tilde{L}_M = \tilde{\phi}_I\frac{\delta L_M}{\delta \phi_I} + (\partial_{\mu}\tilde{\phi}_I)\frac{\delta L_M}{\delta (\partial_{\mu}\phi_I)},
\end{eqnarray}

where $\tilde{\phi}_I = \tilde{\delta}\phi_I$ are the $\tilde{\delta}$ matter fields.   So, the equations of motion are:

\begin{eqnarray}
\label{Einst Eq} G^{\mu \nu} &=& \kappa T^{\mu \nu} \\
\label{tilde Eq} F^{(\mu \nu) (\alpha \beta) \rho
\lambda} D_{\rho} D_{\lambda} \tilde{g}_{\alpha \beta} + \frac{1}{2}g^{\mu \nu}R^{\alpha \beta}\tilde{g}_{\alpha \beta} - \frac{1}{2}\tilde{g}^{\mu \nu}R &=& \kappa\tilde{T}^{\mu \nu},
\end{eqnarray}

with:

\begin{eqnarray}
\label{F}
F^{(\mu \nu) (\alpha \beta) \rho \lambda} &=& P^{((\rho
\mu) (\alpha \beta))}g^{\nu \lambda} + P^{((\rho \nu) (\alpha
\beta))}g^{\mu \lambda} - P^{((\mu \nu) (\alpha \beta))}g^{\rho
\lambda} - P^{((\rho \lambda) (\alpha \beta))}g^{\mu \nu} \nonumber \\
P^{((\alpha \beta)(\mu \nu))} &=& \frac{1}{4}\left(g^{\alpha
\mu}g^{\beta \nu} + g^{\alpha \nu}g^{\beta \mu} - g^{\alpha
\beta}g^{\mu \nu}\right) \nonumber\\
\tilde{T}^{\mu \nu}&=&\tilde{\delta} T^{\mu \nu}, \nonumber
\end{eqnarray}

where $(\mu \nu)$ denotes that $\mu$ and $\nu$ are in a totally symmetric combination. An important fact to notice is that our equations are of second order in derivatives which is needed to preserve causality. We can show  that $(\ref{tilde Eq})_{\mu \nu} = \tilde{\delta}\left[(\ref{Einst Eq})_{\mu \nu}\right]$. Finally, from \textbf{Appendix A}, we have that the action (\ref{grav action}) is invariant under (\ref{trans g}) and (\ref{trans gt}). This means that two conservation rules are satisfied. They are:

\begin{eqnarray}
\label{Conserv T}
D_{\nu}T^{\mu \nu} &=& 0 \\
\label{Conserv tilde T}
D_{\nu}\tilde{T}^{\mu \nu} &=& \frac{1}{2}T^{\alpha \beta}D^{\mu}\tilde{g}_{\alpha \beta} - \frac{1}{2}T^{\mu \beta} D_{\beta}\tilde{g}^{\alpha}_{\alpha} + D_{\beta}(\tilde{g}^{\beta}_{\alpha}T^{\alpha \mu}).
\end{eqnarray}

It is easy to see that (\ref{Conserv tilde T}) is $\tilde{\delta}\left(D_{\nu}T^{\mu \nu}\right) = 0$.

\subsection{\label{SubSec: Test Particle}Test Particle.}

In the previous subsection we found the equations of motion for $\tilde{\delta}$ Gravity.  However, we need to know how the new fields affect the trajectory of a test particle. For this, we will study the test particle action separately for massive and massless particles. The first discussion of this issue in $\tilde{\delta}$ Gravity is in \cite{DG DE}.\\

\subsubsection{\label{SubSubSec: Massive Particles}Massive Particles:}

In GR, the action for a test particle is given by:

\begin{eqnarray}
\label{Geo Action 0}
S_0[\dot{x},g] = - m \int dt \sqrt{-g_{\mu \nu}\dot{x}^{\mu}\dot{x}^{\nu}},
\end{eqnarray}

with $\dot{x}^{\mu} = \frac{d x^{\mu}}{d t}$. This action is invariant under reparametrizations, $t' = t - \epsilon(t)$. In the infinitesimal form is:

\begin{eqnarray}
\label{reparametr}
\delta_R x^{\mu} &=& \dot{x}^{\mu}\epsilon.
\end{eqnarray}

In $\tilde{\delta}$ Gravity, the action is always modified using (\ref{Action}) from \textbf{Appendix A}. So, applying it to (\ref{Geo Action 0}), the new test particle action is:

\begin{eqnarray}
\label{Geo Action 01}
S[\dot{x},y,g,\tilde{g}] &=& - m \int dt \sqrt{-g_{\mu \nu}\dot{x}^{\mu}\dot{x}^{\nu}} + \frac{m}{2} \int dt \left(\frac{\tilde{g}_{\mu \nu}\dot{x}^{\mu}\dot{x}^{\nu}+g_{\mu \nu,\rho}y^{\rho}\dot{x}^{\mu}\dot{x}^{\nu}+2g_{\mu \nu}\dot{x}^{\mu}\dot{y}^{\nu}}{\sqrt{-g_{\mu \nu}\dot{x}^{\mu}\dot{x}^{\nu}}}\right) \nonumber \\
&=& m \int dt \left(\frac{\left(g_{\mu \nu} + \frac{1}{2}\tilde{g}_{\mu \nu}\right)\dot{x}^{\mu}\dot{x}^{\nu} + \frac{1}{2}(2g_{\mu \nu}\dot{y}^{\mu}\dot{x}^{\nu} + g_{\mu \nu, \rho}y^{\rho}\dot{x}^{\mu}\dot{x}^{\nu})}{\sqrt{-g_{\alpha \beta}\dot{x}^{\alpha}\dot{x}^{\beta}}}\right),
\end{eqnarray}

where we have defined $y^{\mu} = \tilde{\delta}x^{\mu}$ and we used that $g_{\mu \nu} = g_{\mu \nu}(x)$, so $\tilde{\delta}g_{\mu \nu} = \tilde{g}_{\mu \nu} + g_{\mu \nu, \rho}y^{\rho}$. Naturally, this action is invariant under reparametrization transformations, given by (\ref{reparametr}), plus $\tilde{\delta}$ reparametrization transformations:

\begin{eqnarray}
\label{reparametr plus}
\delta_R y^{\mu} &=& \dot{y}^{\mu}\epsilon + \dot{x}^{\mu}\tilde{\epsilon},
\end{eqnarray}

just like it is shown in (\ref{tilde trans general}). On the other side, the presence of $y^{\mu}$ suggests additional coordinates, but our model just live in four dimensions, given by $x^{\mu}$. Actually, $y^{\mu}$ can be gauged away using the extra symmetry corresponding to $\tilde{\epsilon}$ in equation (\ref{reparametr plus}), imposing the gauge condition $2g_{\mu \nu}\dot{y}^{\mu}\dot{x}^{\nu} + g_{\mu \nu, \rho}y^{\rho}\dot{x}^{\mu}\dot{x}^{\nu} = 0$. However, the extended general coordinate transformations  (\ref{trans g}) and (\ref{trans gt}), as well as the usual reparametrizations, given by (\ref{reparametr}), are still preserved. Finally, (\ref{Geo Action 01}) is reduced to:

\begin{eqnarray}
\label{Geo Action}
S[\dot{x},g,\tilde{g}] = m \int dt \left(\frac{\left(g_{\mu \nu} + \frac{1}{2}\tilde{g}_{\mu \nu}\right)\dot{x}^{\mu}\dot{x}^{\nu}}{\sqrt{-g_{\alpha \beta}\dot{x}^{\alpha}\dot{x}^{\beta}}}\right).
\end{eqnarray}

Far from the sources, we have the boundary conditions $g_{\mu \nu}\sim \eta_{\mu \nu}$ and $\tilde{g}_{\mu \nu}\sim 0$. In this limit we recover the action for a massive particle  of mass $m$ in Minkowsky space.

This action for a test particle in a gravitational field is the starting point for the physical interpretation of this model. Now, the trajectory of massive test particles is given by the equation of motion of $x^{\mu}$. This equation say us that $g_{\mu \nu}\dot{x}^{\mu}\dot{x}^{\nu} = cte$, just like GR. Now, if we choose $t$ equal to the proper time (See \textbf{Appendix B}), then $g_{\mu \nu}\dot{x}^{\mu}\dot{x}^{\nu} = -1$  and the equation of motion is reduced in this case to:

\begin{eqnarray}
\label{geodesics m}
\hat{g}_{\mu \nu} \ddot{x}^{\nu} + \hat{\Gamma}_{\mu \alpha \beta} \dot{x}^{\alpha} \dot{x}^{\beta} = \frac{1}{4}\tilde{K}_{,\mu},
\end{eqnarray}

with:

\begin{eqnarray}
\hat{\Gamma}_{\mu \alpha \beta} &=& \frac{1}{2}(\hat{g}_{\mu \alpha , \beta} + \hat{g}_{\beta \mu , \alpha} - \hat{g}_{\alpha \beta ,
\mu})\nonumber \\
\hat{g}_{\alpha \beta} &=& \left(1+\frac{1}{2} \tilde{K}\right)g_{\alpha \beta} + \tilde{g}_{\alpha \beta}
\nonumber \\
\tilde{K} &=& \tilde{g}_{\alpha \beta} \dot{x}^{\alpha} \dot{x}^{\beta}. \nonumber
\end{eqnarray}

The equation (\ref{geodesics m}) is a second order equation, but it is not a classical geodesic, because we have additional terms and an effective metric can not be defined. Moreover, the equation of motion is independent of the mass of the particle, so all particles will fall with the same acceleration.\\

\subsection{\label{SubSec: Massless Particles}Massless Particles:}

The massless case is particulary important in this work, because we need to study photon trajectories to define distances. Unfortunately, (\ref{Geo Action 0}) is useless for massless particles, because it is null when $m=0$. To solve this problem, it is a common practice to start from the action \cite{massless geo}:

\begin{eqnarray}
\label{Geo Action 0 Lagr}
S_0[\dot{x},g,v] = \frac{1}{2} \int dt \left(vm^2 - v^{-1}g_{\mu \nu}\dot{x}^{\mu}\dot{x}^{\nu}\right),
\end{eqnarray}

where $v$ is an auxiliary field, which transforms under reparametrizations as:

\begin{eqnarray}
\label{v}
v'(t')&=&\frac{d t}{d t'}v(t).
\end{eqnarray}

From (\ref{Geo Action 0 Lagr}), we can obtain the equation of motion for $v$:

\begin{eqnarray}
\label{v eq}
v = - \frac{\sqrt{-g_{\mu \nu}\dot{x}^{\mu}\dot{x}^{\nu}}}{m}.
\end{eqnarray}

We see from (\ref{v}) that the gauge $v=constant$ can be fixed, so in GR the proper time $\sqrt{-g_{\mu \nu}\dot{x}^{\mu}\dot{x}^{\nu}}$ remains constant along the path. Now, if we substitute (\ref{v eq}) in (\ref{Geo Action 0 Lagr}), we recover (\ref{Geo Action 0}). This means (\ref{Geo Action 0 Lagr}) is equivalent to (\ref{Geo Action 0}), but additionally includes the massless case.\\

In our case, a suitable action, similar to (\ref{Geo Action 0 Lagr}), is:

\begin{eqnarray}
\label{Geo Action Lagr 2}
S[\dot{x},g,\tilde{g},v] = \int dt \left(m^2v - \frac{\left(g_{\mu \nu} +  \tilde{g}_{\mu \nu}\right)\dot{x}^{\mu}\dot{x}^{\nu}}{4v} + \frac{m^2 v^3}{4g_{\alpha \beta}\dot{x}^{\alpha}\dot{x}^{\beta}}\left(m^2 +  v^{-2} \tilde{g}_{\mu \nu}\dot{x}^{\mu}\dot{x}^{\nu}\right)\right).
\end{eqnarray}

In fact, in $\tilde{\delta}$ Gravity the equation of $v$ is still (\ref{v eq}). Thus, by fixing the gauge $v=constant$, the quantity $\sqrt{-g_{\mu \nu}\dot{x}^{\mu}\dot{x}^{\nu}}$ remains constant along the path too. We will use this conserved quantity to define proper time in our model (See \textbf{Appendix B}). Additionally, if we replace (\ref{v eq}) in (\ref{Geo Action Lagr 2}), we obtain the massive test particle action given by (\ref{Geo Action}). But now, we can study the massless case.\\

If we evaluate $m = 0$ in (\ref{Geo Action 0 Lagr}) and (\ref{Geo Action Lagr 2}), we can compare GR and $\tilde{\delta}$ Gravity respectively. They are:

\begin{eqnarray}
\label{Geo Action 0 foton}
S^{(m=0)}_0[\dot{x},g,v] &=& - \frac{1}{2} \int dt v^{-1}g_{\mu \nu}\dot{x}^{\mu}\dot{x}^{\nu} \\
\label{Geo Action foton}
S^{(m=0)}[\dot{x},g,\tilde{g},v] &=& - \frac{1}{4} \int dt v^{-1}\mathbf{g}_{\mu \nu}\dot{x}^{\mu}\dot{x}^{\nu},
\end{eqnarray}

with $\mathbf{g}_{\mu \nu} = g_{\mu \nu} + \tilde{g}_{\mu \nu}$. In both cases, the equation of motion for $v$ implies that a massless particle move in a null-geodesic. In the usual case we have $g_{\mu \nu}\dot{x}^{\mu}\dot{x}^{\nu} = 0$. However, in our model the null-geodesic is given by $\mathbf{g}_{\mu \nu}\dot{x}^{\mu}\dot{x}^{\nu} = 0$, so the trajectory obey a geometry defined by an specifical combination of $g_{\mu \nu}$ and $\tilde{g}_{\mu \nu}$, $\mathbf{g}_{\mu \nu} = g_{\mu \nu} + \tilde{g}_{\mu \nu}$. The equation of motion for the path of a test massless particle is given by:

\begin{eqnarray}
\label{geodesics null}
\mathbf{g}_{\mu \nu} \ddot{x}^{\nu} + \mathbf{\Gamma}_{\mu \alpha \beta} \dot{x}^{\alpha} \dot{x}^{\beta} = 0 \\
\mathbf{g}_{\mu \nu}\dot{x}^{\mu}\dot{x}^{\nu} = 0, \nonumber
\end{eqnarray}

with:

\begin{eqnarray}
\mathbf{\Gamma}_{\mu \alpha \beta} = \frac{1}{2}(\mathbf{g}_{\mu \alpha , \beta} + \mathbf{g}_{\beta \mu , \alpha} - \mathbf{g}_{\alpha \beta, \mu}). \nonumber
\end{eqnarray}

In \textbf{Appendix B}, the proper time defined for massive particles and the effective metric of massless particles are used to study the cosmological geometry. With this, we will define the effective scale factor to explain the accelerated expansion of the universe without a cosmological constant. This calculation will be elaborated in \textbf{Section \ref{Sec: Cosmological Case}}.\\

To summarize, we obtained the equations of motion of $\tilde{\delta}$ Gravity, given by (\ref{Einst Eq}), (\ref{tilde Eq}), (\ref{Conserv T}) and (\ref{Conserv tilde T}). In \textbf{Appendix C}, are presented the energy momentum tensors for a perfect fluid in equations (\ref{T PF}) and (\ref{T tilde PF}). Then, we obtained how a test particle moves when it is coupled to $g_{\mu \nu}$ and $\tilde{g}_{\mu \nu}$, given by (\ref{geodesics m}) or (\ref{geodesics null}) if we have a massive or massless particle respectively. In the next section  we will obtain the accelerated expansion of the universe assuming a universe without dark energy.\\


\section{\label{Sec: Cosmological Case}Cosmological Case.}

In this section we will study photons emitted from a supernova using $\tilde{\delta}$ Gravity to explain the accelerated expansion of the universe without dark energy. For this, we have to use the correct cosmological geometry to represent an homogeneous and isotropic universe, given by the FLRW metric. In the harmonic coordinate system it is:

\begin{eqnarray}
g_{\mu \nu}dx^{\mu}dx^{\nu} =  - T^2(u)c^2du^2 + R^2(u)\left(dx^2 + dy^2 + dz^2\right), \nonumber
\end{eqnarray}

such that $T(u) = \frac{d t}{d u}(u)$ and $t$ is the cosmological time. In the same form, $\tilde{g}_{\mu \nu}$ is given by:

\begin{eqnarray}
\tilde{g}_{\mu \nu}dx^{\mu}dx^{\nu} =  - F_b(u)T^2(u)c^2du^2 + F_a(u)R^2(u)\left(dx^2 + dy^2 + dz^2\right). \nonumber
\end{eqnarray}

Now, if we impose (\ref{Harmonic gauge}) and (\ref{Harmonic gauge tilde}) from \textbf{Appendix D} to fix the harmonic gauge, we obtain that $T(u) = T_0 R^3(u)$ and $F_b(u) = 3(F_a(u) + T_1)$, where $T_0$ and $T_1$ are gauge constants. We use $T_0 = 1$ and $T_1 = 0$ to fix the gauge completely. So, with these conditions, the system ($u$,$x$,$y$,$z$) correspond to harmonic coordinate. Now, we can return to the usual system where $g_{\mu \nu}$ and $\tilde{g}_{\mu \nu}$ are given by:\\

\begin{eqnarray}
\label{g FLRW}
g_{\mu \nu}dx^{\mu}dx^{\nu} &=&  - c^2 dt^2 + R^2(t)\left(dx^2 + dy^2 + dz^2\right) \\
\label{gt FLRW}
\tilde{g}_{\mu \nu}dx^{\mu}dx^{\nu} &=&  - 3F_a(t) c^2 dt^2 + F_a(t)R^2(t)\left(dx^2 + dy^2 + dz^2\right).
\end{eqnarray}

These expressions represent an isotropic and homogeneous universe. From \textbf{Appendix B}, we know that the proper time is measured only using the metric $g_{\mu \nu}$, but the space geometry is determined by the modified null-geodesic, given by (\ref{geodesics null}), where both tensor fields, $g_{\mu \nu}$ and $\tilde{g}_{\mu \nu}$, are needed. These considerations are fundamental to explain the expansion of the universe with $\tilde{\delta}$ Gravity. Now, with all these, we can study a photon trajectory from a supernova and solve the equations of motion.\\

\subsection{\label{SubSec:Photon Trajectory and Luminosity Distance}Photon Trajectory and Luminosity Distance:}

In this section we follow \cite{DG DE}. When a photon emitted from a supernova travels to the Earth, the Universe is expanding. This means that the photon is affected by the cosmological Doppler effect. For this, we must use a null geodesic, given by (\ref{geodesics null}), in a radial trajectory from $r_1$ to $r=0$. Therefore, using (\ref{g FLRW}) and (\ref{gt FLRW}), we obtain:

\begin{eqnarray}
- (1+3 F_a(t)) c^2 dt^2 + R^2(t)(1+ F_a(t)) dr^2 = 0. \nonumber
\end{eqnarray}

Define the effective scale factor (See \textbf{Appendix B}):

\begin{eqnarray}
\label{R tilde}
\tilde{R}(t) = R(t)\sqrt{\frac{1+ F_a(t)}{1+3 F_a(t)}}
\end{eqnarray}

such that $c dt = - \tilde{R}(t) dr$. Now, if we integrate this expression from $r_1$ to $0$, we obtain:

\begin{eqnarray}
\label{r1 1}
r_1 = c \int_{t_1}^{t_0} \frac{dt}{\tilde{R}(t)},
\end{eqnarray}

where $t_1$ and $t_0$ are the emission and reception times. If a second wave crest is emitted at $t = t_1 + \Delta t_1$ from $r = r_1$, it will reach $r = 0$ at $t = t_0 + \Delta t_0$, so:

\begin{eqnarray}
\label{r1 2}
r_1 = c \int_{t_1 + \Delta t_1}^{t_0 + \Delta t_0} \frac{dt}{\tilde{R}(t)}.
\end{eqnarray}

Therefore, if $\Delta t_0$ and $\Delta t_1$ are small, which is appropriate for light waves, we get:

\begin{eqnarray}
\frac{\Delta t_0}{\Delta t_1} = \frac{\tilde{R}(t_0)}{\tilde{R}(t_1)}
\end{eqnarray}

Since $t$ measures proper time, we get:

\begin{eqnarray}
\frac{\Delta \nu_1}{\Delta \nu_0} = \frac{\tilde{R}(t_0)}{\tilde{R}(t_1)},
\end{eqnarray}

where $\nu_0$ is the light frequency detected at $r = 0$, corresponding to a source emission at frequency $\nu_1$. So, the redshift is given by:

\begin{eqnarray}
\label{redshift}
1 + z(t_1) = \frac{\tilde{R}\left(t_0\right)}{\tilde{R}(t_1)}.
\end{eqnarray}

We see that $\tilde{R}\left(t\right)$ replaces the usual scale factor $R(t)$ to compute $z$. This means that we need to redefine the luminosity distance too. For this, let us consider a mirror of radius $b$ that is receiving light from our distant source at $r_1$. The photons that reach the mirror are within a cone of half-angle $\epsilon$ with origin at the source.\\

Let us compute $\epsilon$. The path of the light rays is given by $\vec{r}(\rho) = \rho \hat{n} + \vec{r}_1$, where $\rho > 0$ is a parameter and $\hat{n}$ is the direction of the light ray. Since the mirror is in $\vec{r} = 0$, then $\rho = r_1$ and $\hat{n} = - \hat{r}_1 + \vec{\epsilon}$, where $\epsilon$ is the angle between $-\vec{r}_1$ and $\hat{n}$ at the source, forming a cone. The proper distance is determined by the tri-dimensional metric, given by (See \textbf{Appendix B}):

\begin{eqnarray}
d l^2 &=& \gamma_{ij}dx^{i}dx^{j} \nonumber \\
&=& \tilde{R}^2(t)\delta_{i j}dx^{i}dx^{j} \nonumber
\end{eqnarray}

in the cosmological case. Then $b = \tilde{R}(t_0) r_1 \epsilon$ and the solid angle of the cone is:

\begin{eqnarray}
\Delta \Omega &=& \int_0^{2 \pi} d\phi \int_0^{\epsilon} \sin(\theta) d\theta = 2\pi (1-\cos(\epsilon)) \nonumber \\
&=& \pi \epsilon^2 = \frac{A}{r_1^2 \tilde{R}^2(t_0)}, \nonumber
\end{eqnarray}

where $A = \pi b^2$ is the proper area of the mirror. This means that $\epsilon = \frac{b}{r_1 \tilde{R}(t_0)}$. So, the fraction of all isotropically emitted photons that reach the mirror is:

\begin{eqnarray}
f &=& \frac{\Delta \Omega}{4 \pi} \nonumber \\
&=& \frac{A}{4 \pi r_1^2 \tilde{R}^2(t_0)}. \nonumber
\end{eqnarray}

We know that the apparent luminosity, $l$, is the received power per unit mirror area. Power is energy per unit time, so the received power is $P = \frac{h\nu_0}{\Delta t_0} f$, where $h\nu_0$ is the energy corresponding to the received photon. On the other side, the total emitted power by the source is $L = \frac{h\nu_1}{\Delta t_1}$, where $h\nu_1$ is the energy corresponding to the emitted photon. Therefore, we have that:

\begin{eqnarray}
P &=& \frac{\tilde{R}^2(t_1)}{\tilde{R}^2(t_0)} L f \nonumber \\
l &=& \frac{P}{A} \nonumber \\
&=& \frac{\tilde{R}^2(t_1)}{\tilde{R}^2(t_0)} \frac{L}{4 \pi r_1^2 \tilde{R}^2(t_0)}, \nonumber
\end{eqnarray}

where we have used that $\frac{\Delta t_0}{\Delta t_1} = \frac{\nu_1}{\nu_0} = \frac{\tilde{R}(t_0)}{\tilde{R}(t_1)}$. Besides, we know that, in an Euclidean space, the luminosity decreases with distance $d_L$ according to $l = \frac{L}{4\pi d_L^2}$. Therefore, using (\ref{r1 1}), the luminosity distance is:

\begin{eqnarray}
\label{d_L 00}
d_L &=& \frac{\tilde{R}^2(t_0)}{\tilde{R}(t_1)} r_1 \nonumber \\
&=& c \frac{\tilde{R}^2\left(t_0\right)}{\tilde{R}(t_1)} \int_{t_1}^{t_0} \frac{dt}{\tilde{R}(t)}.
\end{eqnarray}

We can also define the angular diameter distance, $d_A$\footnote{We follow the discussion in \cite{kolb}}. Let us consider a source with proper size $D$, situated at $r=r_1$, that emits photons at $t=t_1$. These photons reach us($r=0$) at $t=t_0$. From (\ref{tri metric}) in \textbf{Appendix B}, the observed angular diameter of the source is $\theta=\frac{D}{\tilde{R}(t_1)r_1}$. The angular diameter distance is defined by $d_A = \frac{D}{\theta}$, so $d_A = \tilde{R}(t_1)r_1$. If we compare it with (\ref{d_L 00}), we obtain that:

\begin{eqnarray}
\label{d_A d_L}
d_A
&=& \frac{\tilde{R}^2(t_1)}{\tilde{R}^2(t_0)}d_L \nonumber \\
&=& \frac{d_L}{(1+z_1)^2}.
\end{eqnarray}

Therefore, the relation between $d_A$ and $d_L$ is the same as in GR \cite{weinberg cosmo}. This result is important, because in other modified gravity theories this relation is not satisfied \cite{dA dL}. We will use $d_L$ to analyze the supernovae data, but $d_A$ could be useful for other phenomena. In the next sections, we will solve the equations of motion and fit the supernovae data.\\

\subsection{\label{SubSec: Equations Solution}Equations Solution:}

In cosmology, the metric $g_{\mu \nu}$ is given by (\ref{g FLRW}). Besides, by (\ref{Einst Eq}) and (\ref{Conserv T}), we know that Einstein's equations do not change and $T_{\mu \nu}$ is conserved. Therefore, the usual cosmological solution is still valid. So, using (\ref{T PF}) from \textbf{Appendix C} with $U_{\mu} = (c,0,0,0)$, we obtain the well-known equations:

\begin{eqnarray}
\label{Eq R(t)}
\left(\frac{\dot{R}(t)}{R(t)}\right)^2 &=& \frac{\kappa c^2}{3} \sum_i \rho_i(t) \\
\label{Eq rho(t)}
\dot{\rho}_i(t) &=& - \frac{3\dot{R}(t)}{R(t)}(\rho_i(t) + p_i(t)),
\end{eqnarray}

with $\dot{f}(t) = \frac{d f}{d t}(t)$ and we assumed that the interaction between different components of the universe is null. Additionally, to solve (\ref{Eq R(t)}) and (\ref{Eq rho(t)}), we need equations of state which relate $\rho_i(t)$ and $p_i(t)$, for which we take $p_i(t) = \omega_i\rho_i(t)$. Since we wish to explain dark energy with $\tilde{\delta}$ Gravity, we will assume that in the Universe we only have non-relativistic matter (cold dark matter, baryonic matter) and radiation (photons, massless particles). So, we will require two equations of state. For non-relativistic matter we use $p_M(t) = 0$ and for radiation $p_R(t) = \frac{1}{3}\rho_R(t)$. Replacing in (\ref{Eq R(t)}) and (\ref{Eq rho(t)}) and solving them, we find the exact solution:

\begin{eqnarray}
\rho(Y) &=& \rho_M(Y) + \rho_R(Y) \nonumber \\
\label{FLRW Sol 1}&=& \frac{3H_0^2\Omega_R}{\kappa c^2 C} \frac{Y + C}{Y^4} \\
p(Y) &=& \frac{1}{3} \rho_R(t) \nonumber \\
\label{FLRW Sol 2}&=& \frac{H_0^2\Omega_R}{\kappa c^2} \frac{1}{Y^4} \\
\label{FLRW Sol 3}
t(Y) &=& \frac{2\sqrt{C}}{3H_0\sqrt{\Omega_R}}\left(\sqrt{Y+C}(Y-2C) + 2C^{\frac{3}{2}}\right) \\
Y &=& \frac{R(t)}{R_0},
\end{eqnarray}

where $t(Y)$ is the time variable, $R_0$ is the scale factor in the present, $C = \frac{\Omega_R}{\Omega_M}$, and $\Omega_R$ and $\Omega_M = 1 - \Omega_R$ are the radiation and non-relativistic matter density in the present respectively. We know that $\Omega_R \ll 1$, so $\Omega_M \sim 1$ and $C \ll 1$. We can see that it is convenient to use $Y$ like our independent variable. By definition, $Y \gg C$ describes the non-relativistic era and $Y \ll C$ describes the radiation era.\\

The equation of motion for $\tilde{g}_{\mu \nu}$ is given by (\ref{tilde Eq}) and (\ref{Conserv tilde T}), where $\tilde{T}_{\mu \nu}$ is a new energy-momentum tensor for $\tilde{\delta}$ non-relativistic matter and radiation densities, given by $\tilde{\rho}_M$ and $\tilde{\rho}_R$ respectively. So, using (\ref{FLRW Sol 1})-(\ref{FLRW Sol 3}) and (\ref{T tilde PF}) from \textbf{Appendix C} and $Y$ as the independent variable, (\ref{tilde Eq}) and (\ref{Conserv tilde T}) are reduced to:

\begin{eqnarray}
\label{tEQ FLRW 1}
F_a'(Y) + \frac{2c\kappa}{9H_0^2\Omega_M}Y^2\left(Y\tilde{\rho}_M'(Y) + 3\tilde{\rho}_M(Y)\right) &=& 0 \\
\label{tEQ FLRW 2}
F_a'(Y) + \frac{c\kappa}{6H_0^2\Omega_R}Y^3\left(Y\tilde{\rho}_R'(Y) + 4\tilde{\rho}_R(Y)\right) &=& 0 \\
\label{tEQ FLRW 3}
YF_a'(Y) - 3F_a(Y) - \frac{c\kappa}{3H_0^2\Omega_M}\frac{Y^4}{\left(C+Y\right)}\left(\tilde{\rho}_M(Y)+\tilde{\rho}_R(Y)\right) &=& 0,
\end{eqnarray}

where we used $\tilde{p}_M = 0$, $\tilde{p}_R = \frac{1}{3}\tilde{\rho}_R$ and $U^T_{\mu} = 0$. The solution of these equations are:

\begin{eqnarray}
\label{FLRW Sol 4}
F_a(Y) &=& \frac{3}{2}\left(2C_2-C_1\right)\frac{Y}{C}\left(\sqrt{\frac{Y}{C}+1}\ln\left(\frac{\sqrt{\frac{Y}{C}+1}+1}{\sqrt{\frac{Y}{C}+1}-1}\right) - 2\right) \nonumber \\ && - 2C_2 + C_3\frac{Y}{C}\sqrt{\frac{Y}{C}+1} \\
\label{FLRW Sol 5}
\tilde{\rho}_M(Y) &=& \frac{9H_0^2\Omega_R}{2\kappa c^2 C} \frac{(C_1-F_a(Y))}{Y^3} \\
\label{FLRW Sol 6}
\tilde{\rho}_R(Y) &=& \frac{6H_0^2\Omega_R}{\kappa c^2} \frac{(C_2-F_a(Y))}{Y^4},
\end{eqnarray}

where $C_1$, $C_2$ and $C_3$ are integration constants. $\tilde{\rho}_M(Y)$ and $\tilde{\rho}_R(Y)$ are densities of $\tilde{\delta}$ matter, so they must be not-negative functions. Then:

\begin{eqnarray}
\label{pos}
C_1 - F_{a}(Y) \geq 0 &\wedge& C_2 - F_{a}(Y) \geq 0 \textrm{, for all } Y \geq 0.
\end{eqnarray}

Evaluating (\ref{pos}) at $Y=0$, we get $C_2 \geq 0$ and $2C_{2} + C_{1} \geq 0$. On the other side, at $Y \gg C$, we get $C_{3} \leq 0$.\\

On the other side, if we use (\ref{FLRW Sol 4}) in (\ref{R tilde}) and define $\tilde{Y} = \frac{\tilde{R}(t)}{R\left(t_0\right)}$, we can see that:

\begin{eqnarray}
\frac{\tilde{Y}}{Y} \sim \sqrt{\frac{1-2C_2}{1-6C_2}}\left(1 + \frac{3\left(2C_2 - C_1\right)}{2C\left(1-6C_2\right)\left(1-2C_2\right)}Y\log(Y)\right) + O(Y),
\end{eqnarray}

when $Y \ll C$. $\tilde{Y}$ is the effective scale factor, so represent the evolution of the universe. We know that an accelerated expansion must be produced at late times, but the expansion must be driven by the non-relativistic matter and radiation at early times, this means $\frac{\tilde{Y}}{Y} = 1 + O(Y)$. For this, we have to fix $C_1 = 0$ and $C_2=0$ to guarantee the temporal behavior of expansion is just like GR at early times. The other constants will be chosen such that a Big-Rip is produced. That is $\tilde{Y}(Y_{Rip}) = \infty$. We need a Big-Rip to explain the accelerated expansion of the universe because we want that $\tilde{Y}$ to grow quickly when $Y$ is bigger. In \textbf{Appendix B}, we proved that the tri-dimensional metric is positive definite until the Big-Rip is produced.\\

The Big-Rip is determined by $C_3$, but it is necessary a very small value for this parameter, if not the Big-Rip would be too early. However, if we use $C_3 = 0$, the Big-Rip is not produced and we cannot explain the accelerated expansion of the universe. So, using $C_3 = - \frac{C^{^\frac{3}{2}}L_2}{3}$,with  $1 \gg C \neq 0$, the effective scale factor is given by:

\begin{eqnarray}
\label{Mod Scale Factor FLRW}
\tilde{Y}(Y,L_1,L_2,C)
= Y\sqrt{\frac{1-L_2\frac{Y}{3}\sqrt{Y+C}}{1-L_2Y\sqrt{Y+C}}}.
\end{eqnarray}

 From (\ref{Mod Scale Factor FLRW}), it is clear that the Big-Rip is produced when:

\begin{eqnarray}
\label{Y rip}
Y_{Rip} = \left(\frac{1}{L_2}\right)^{\frac{2}{3}}.
\end{eqnarray}

In summary, we have that $\tilde{Y} \sim Y$ in the radiation era, where $Y \ll C$, so the Universe evolves without differences with GR. However, in some moment during the non-relativistic matter era, where $Y \gg C$, an accelerated expansion is produced, ending in a Big-Rip. We will give more details for this when we study the supernovae data. Additionally, inequalities in (\ref{pos}) are always satisfied, then the $\tilde{\delta}$ densities are non-negative.\\

\subsection{\label{SubSec: Analysis and Results}Analysis and Results:}

Before we start the data analysis, we must define the parameters of the model. In the  first place, $d_L$ in GR depends upon four parameters: $Y$, $H_0 = 100 h$ km s$^{-1}$ Mpc$^{-1}$, $\Omega_M$ and $\Omega_R$. However, from CMB black body spectrum we obtain the photons density in the present, $\Omega_{\gamma}$. Now, if we assume that $\Omega_R = \Omega_{\gamma} + \Omega_{\nu} = \left(1+3 \left(\frac{7}{8}\right)\left(\frac{4}{11}\right)^{4/3}\right) \Omega_{\gamma}$($\Omega_{\nu}$is the primordial neutrino density), we get $h^2\Omega_R = 4.15 \times 10^{-5}$. Therefore, the parameters in $d_L$ can be reduced to three: $Y$, $h$ and $h^2\Omega_M$. In the same way, in $\tilde{\delta}$ Gravity with $\tilde{\delta}$ matter, $d_L$ depends on three parameters: $Y$, $C$ and $L_2$. We will use $H_0 \sqrt{\Omega_R} = 0.644$ km s$^{-1}$ Mpc$^{-1}$.\\

The supernovae data gives the apparent magnitude, $m$, as a function of redshift, $z$. For this reason, it is useful to use $z$ instead of $Y$. The apparent magnitude is:

\begin{eqnarray}
\label{m}
m(z) = M + 5\log_{10}\left(\frac{d_L(z)}{10 \textrm{ pc}}\right),
\end{eqnarray}

where $M$ is the absolute magnitude, constant and common for all supernovae. The difference between GR and $\tilde{\delta}$ Gravity is in $d_L(z)$. In GR, we have that:

\begin{eqnarray}
\label{d_L usual}
d_L(z) =  \frac{c (1+z) \textrm{ Mpc s}}{100 \textrm{ km}} \int_{\frac{1}{1 + z}}^{1} \frac{dY'}{\sqrt{h^2\Omega_{\Lambda} Y'^4 + h^2\Omega_M Y' + h^2\Omega_R}},
\end{eqnarray}

with $\Omega_{\Lambda} = 1 - \Omega_M - \Omega_R$. From (\ref{d_L usual}), we can fit the data for GR. On the other side, in our modified gravity model, the luminosity distance is given by (\ref{d_L 00}). So, using the relation in (\ref{FLRW Sol 3}), we have, for $\tilde{\delta}$ Gravity, that:

\begin{eqnarray}
\label{d_L}
d_L(z) &=&  c (1+z)\frac{\sqrt{C}}{H_0\sqrt{\Omega_R}}\int_{0}^{z} \frac{(1+u)Y(u)Y'(u)}{\sqrt{Y(u)+C}} du,
\end{eqnarray}

where $Y'(z) = \frac{dY}{dz}(z)$. Besides, to find $Y(z)$ we must solve (\ref{redshift}). That is:

\begin{eqnarray}
\label{z}
\tilde{Y}(Y(z)) &=& \frac{\tilde{Y}_0}{1 + z},
\end{eqnarray}

where $\tilde{Y}_0 = \tilde{Y}(1)$ and $\tilde{Y}(Y(z))$ is given by (\ref{Mod Scale Factor FLRW}).\\

The statistical method used to interpret errors in data is given by the variance $\sigma$ in a normally distributed random variable. This means, if we are fitting a function $y(x)$ with a set of points $(x_i, y_i)$, we must minimize \cite{Numeric}:

\begin{eqnarray}
\chi^2(\textrm{per point}) &=& \frac{1}{N}\sum^N_{i=1} \frac{\left(y_i - y (x_i)\right)^2}{\sigma_{y_i}^2}, \nonumber
\end{eqnarray}

where $N$ is the number of data points and $\sigma_{y_i}$ is the error of $y_i$. In our case, the data is given by $(z_i, \mu_i)$, where:

\begin{eqnarray}
\mu(z) \equiv m(z) - M = 5 \log_{10} \left(\frac{d_L(z)}{10\textrm{ pc}}\right) \nonumber
\end{eqnarray}

is the distance modulus. Then:

\begin{eqnarray}
\label{chi square}
\chi^2(\textrm{per point}) = \frac{1}{N}\sum^N_{i=1} \frac{\left(\mu_i - \mu(z_i)\right)^2}{\sigma_{\mu_i}^2}.
\end{eqnarray}

Now, we can proceed to analyze the data given in \cite{Data} with $N = 580$ supernovae. In both cases, GR and $\tilde{\delta}$ Gravity, $d_L$ is given by an exact expression, but we need to use a numerical method to solve the integral and fit the data to determinate the optimum values for the parameters that represent the $\mu$ v/s $z$ of the supernovae data. For this, we used mathematica 11.0 \footnote{To obtain the best combination of parameters, we used \textbf{NonLinearModelFit}. Then, we used them to minimize (\ref{chi square}). See the Mathematica 11.0 help for more details.}. The parameters that minimize (\ref{chi square}) are:\\

In GR: $h = 0.7 \pm 3.37 \times 10^{-3}$ and $h^2\Omega_M = 0.136 \pm 8.5 \times 10^{-3}$ with $\chi^2(\textrm{per point}) = 0.985$.\\

In $\tilde{\delta}$ Gravity with $\tilde{\delta}$ matter: $L_2 = 0.457 \pm 0.0114$ and $C = 1.89 \times 10^{-4} \pm 4.92 \times 10^{-6}$ with $\chi^2(\textrm{per point}) = 0.985$.\\

\begin{figure}
\begin{center}
\includegraphics[scale=0.85]{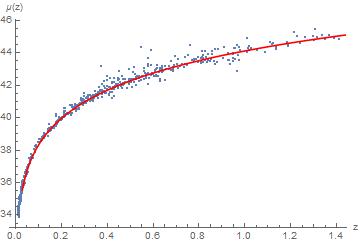}
\caption{{\scriptsize Distance modulus vs Redshift. We have fitted $580$ supernovae to $\tilde{\delta}$ Gravity, using the parameters that minimize (\ref{chi square}).}}
\label{Fig: Fit}
\end{center}
\end{figure}

We can see in \textbf{Figure \ref{Fig: Fit}} that $\tilde{\delta}$ Gravity with $\tilde{\delta}$ matter fit the data very well. Now, with these values, we can compute the age of the universe and the Big-Rip era. For GR, the age of the universe is $1.377 \times 10^{10}$ years. However, in our model, the time is given by (\ref{FLRW Sol 3}). So, substituting the corresponding values for $L_2$, $C$ and taking $Y = \frac{R(t)}{R(t_0)} = 1$, we obtain $1.391\times 10^{10}$ years for the age of the universe. To compute when the Big-Rip will happen, we need to use (\ref{Y rip}). That is $Y_{Rip} = 1.684$, so $t_{\textrm{Big-Rip}} = 3.042 \times 10^{10}$ years. Therefore, the Universe has lived less than half of its life.\\

On the other side, in \cite{Paper DE} we obtained that, in $\tilde{\delta}$ Gravity without $\tilde{\delta}$ matter, the age of the universe is $1.92 \times 10^{10}$ years and $t_{\textrm{Big-Rip}} = 4.3 \times 10^{10}$ years. The problem in this case is the huge age of the universe, compared with Planck Collaboration given by $1.381 \times 10^{10}$ years\footnote{The age of the universe of Planck was calculated using the cosmological parameters obtained in \cite{planck age}. That is $\Omega_M = 0.308$ and $H_0 \equiv 100h = 67.8$ km/s/Mpc.}. However, we cannot say that this case is totally rejected yet, but the age of the universe for $\tilde{\delta}$ Gravity with $\tilde{\delta}$ matter is more similar to Planck.\\

When we obtained the effective scale factor, given by (\ref{Mod Scale Factor FLRW}), it was clear that it is not possible to obtain a Big-Rip if $C = 0$. This means that, in $\tilde{\delta}$ Gravity, a minimal component of radiation is necessary obtain an accelerated expansion of the Universe and explain the supernovae data without dark energy, so $1 \gg C \neq 0$. In this way, the accelerated expansion of the Universe can be understood as a geometric effect.\\

On the other side, using (\ref{FLRW Sol 5}-\ref{FLRW Sol 6}), the $\tilde{\delta}$ matter in the present is given by:

\begin{eqnarray}
\tilde{\rho}_{M0} \equiv \tilde{\rho}_M(1) &\approx& \frac{3H_0^2\Omega_R L_2}{2\kappa c^2 C} \\
\tilde{\rho}_{R0} \equiv \tilde{\rho}_R(1) &\approx& \frac{2H_0^2\Omega_R L_2}{\kappa c^2}.
\end{eqnarray}

So, defining the normalize $\tilde{\delta}$ densities in the present:

\begin{eqnarray}
\tilde{\Omega}_M &=& \frac{\kappa c^2\tilde{\rho}_{M0}}{3H_0^2} \nonumber \\
\tilde{\Omega}_R &=& \frac{\kappa c^2\tilde{\rho}_{R0}}{3H_0^2}, \nonumber
\end{eqnarray}

we obtain:

\begin{eqnarray}
\tilde{\Omega}_M &\approx& \frac{L_2}{2}\Omega_M \nonumber \\
&\approx& 0.23\Omega_M \\
\tilde{\Omega}_R &\approx& \frac{2L_2}{3}\Omega_R \nonumber \\
&\approx& 0.3\Omega_R.
\end{eqnarray}

Therefore, we have two components of $\tilde{\delta}$ matter, related to ordinary matter at cosmological level. These components can be considered like a contribution to Dark Matter, however a more accuracy analysis in a field theory level is necessary to understand the nature of $\tilde{\delta}$ matter. In any case, $\tilde{\Omega}_M$ and $\tilde{\Omega}_R$ are important to explain the dynamic of the expansion of the universe. So, the observation of others cosmological phenomena, like the CMB, must be taken into account.\\

In conclusion, we have two points of view of $\tilde{\delta}$ Gravity. In the first one, we do not have $\tilde{\delta}$ matter and the new symmetry is broken. This computation is explained with more details in \cite{Paper DE}. In the second one, we have $\tilde{\delta}$ matter, so the new symmetry is preserved. For this computation, we have used an action to describe a perfect fluid, given by (\ref{Action PF}). To preserve the new symmetry, it is necessary a new energy-momentum tensor, $\tilde{T}_{\mu \nu}$, given by the equation (\ref{Conserv tilde T}). To compute this tensor, we need to use an off-shell expression for $T_{\mu \nu}$, unlike in the first case, where we used an on-shell expression. All of these differences give us different results, but the same effect, an accelerated expansion of the universe without dark energy. $\tilde{\delta}$ Gravity is a recent work, therefore we cannot discard any of them. We must consider both cases like two different versions of $\tilde{\delta}$ Gravity.\\

To conclude the cosmological case, we will briefly comment about the general equations of motion for only one fluid in $\tilde{\delta}$ Gravity for future references. Using the cosmological solution for $g_{\mu \nu}$ and $\tilde{g}_{\mu \nu}$, given by (\ref{g FLRW}) and (\ref{gt FLRW}) respectively, we obtain:

\begin{eqnarray}
\label{eq1}
\kappa \rho(t) &=& 3H^2(t) \\
\label{eq2}
\kappa p(t) &=& - 2\dot{H}(t) - 3H^2(t) \\
\label{eqt1}
\kappa \tilde{\rho}(t) &=& 3H(t)\left(\dot{F}_a(t)-3H(t)F_a(t)\right) \\
\label{eqt2}
\kappa \tilde{p}(t) &=& - \ddot{F}_a(t) + 3\left(2\dot{H}(t) + 3H^2(t)\right)F_a(t),
\end{eqnarray}

where $H(t) = \frac{\dot{R}(t)}{R(t)}$ is the Hubble parameter and $\tilde{p}(t) = \tilde{\delta}p(t)$. To complete the system, we need equations of state to solve them. They are $p(t) = \omega(t)\rho(t)$ and $\tilde{p}(t) = \omega(t)\tilde{\rho}(t) + \tilde{\omega}(t)\rho(t)$. In a perfect fluid, $\omega(t)$ usually is assumed to be constant and $\tilde{\omega}(t)$ must be zero in that case. So, using these equations of state in (\ref{eq1}-\ref{eqt2}), we obtain:

\begin{eqnarray}
\label{EQ 1}
\dot{H}(t) + \frac{3}{2}\left(\omega(t)+1\right)H^2(t) &=& 0 \\
\label{EQ 2}
\ddot{F}_a(t) + 3\omega(t)H(t)\dot{F}_a(t) &=& - 3\tilde{\omega}(t)H^2(t).
\end{eqnarray}

In order to understand the behavior of these equations, we will solve the case where $\omega(t)$ is almost constant, to be close to a  perfect fluid. The solution is:

\begin{eqnarray}
\label{w cte}
R(t) &=&
\left\{
  \begin{array}{cc}
    R_0\left(t+t_0\right)^{\frac{2}{3\left(1+\omega\right)}} & \textrm{, with: } \omega \neq - 1 \\
    R_1e^{\lambda t} & \textrm{, with: } \omega = - 1 \\
  \end{array}
\right. \nonumber \\
F_a(t) &=&
\left\{
  \begin{array}{cc}
    a_0\left(\frac{R(t)}{R_2}\right)^{\frac{3}{2}\left(1-\omega\right)} + \frac{2\tilde{\omega}}{1-\omega}\ln\left(\frac{R(t)}{R_2}\right) & \textrm{, with: } \omega \neq 1 \\
    a_1\ln\left(\frac{R(t)}{R_3}\right) - \frac{3\tilde{\omega}}{2}\ln^2\left(\frac{R(t)}{R_3}\right) & \textrm{, with: } \omega = 1 \\
  \end{array}
\right.,
\end{eqnarray}

where $R_0$, $R_1$, $R_2$, $R_3$, $t_0$, $a_0$ and $a_1$ are integration constants. From this result, we notice that $R(t)$ obey the standard power-law solution in a perfect fluid with a constant $\omega$. However, we must remember that the dynamic of the universe in $\tilde{\delta}$ Gravity is given by the effective scale factor, (\ref{R tilde}), that produce an accelerated expansion. This means that we can define an effective Hubble parameter given by $\mathcal{H}(t) \equiv \frac{\dot{\tilde{R}}(t)}{\tilde{R}(t)}$. In this case, with $\tilde{\omega} = 0$, it is:

\begin{eqnarray}
\label{Hubble eff}
\mathcal{H}(X) &=&
\left\{
  \begin{array}{cc}
    H(X)\left(1 + \frac{3c_2\left(1 - \omega\right)X^{\frac{3}{2}\left(1-\omega\right)}}{2\left(c_1+2c_2-X^{\frac{3}{2}\left(1-\omega\right)}\right)\left(c_1-X^{\frac{3}{2}\left(1-\omega\right)}\right)}\right) & \textrm{, with: } \omega \neq 1 \\
    & \\
    H(X)\left(1 + \frac{d_2}{\left(d_1+2d_2-\ln(X)\right)\left(d_1-\ln(X)\right)}\right) & \textrm{, with: } \omega = 1, \\
  \end{array}
\right.
\end{eqnarray}

where $X = \frac{R(t)}{R_0}$ and $c_1$, $c_2$, $d_1$ and $d_2$ are integration constant. Therefore, even with a standard power-law solution, we can obtain a different behavior. However, we have to say that the power-law solution is just for a perfect fluid. Actually, the solution of $R(t)$ in some specific model will be the same solution obtained in GR. That is because the Einstein's equation are preserved in $\tilde{\delta}$ Gravity, but the dynamic is affected by the effective scale factor. For example, in inflation, a scalar field is used to produce the exponential expansion. In that case:

\begin{eqnarray}
\rho(t) &=& \frac{1}{2}\dot{\varphi}_0^2(t) + V(\varphi_0(t)) \nonumber \\
p(t) &=& \frac{1}{2}\dot{\varphi}_0^2(t) - V(\varphi_0(t)) \nonumber \\
\tilde{\rho}(t) &=& \dot{\varphi}_0(t)\dot{\tilde{\varphi}}_0(t) - \frac{3}{2}F_a(t)\dot{\varphi}_0^2(t) + V_{,\varphi}(\varphi_0(t))\tilde{\varphi}_0(t) \nonumber \\
\tilde{p}(t) &=& \dot{\varphi}_0(t)\dot{\tilde{\varphi}}_0(t) - \frac{3}{2}F_a(t)\dot{\varphi}_0^2(t) - V_{,\varphi}(\varphi_0(t))\tilde{\varphi}_0(t). \nonumber
\end{eqnarray}

With GR, inflation must obey $V(\varphi_0(t)) \gg \dot{\varphi}_0^2(t)$ to obtain $\omega(t) = \frac{p(t)}{\rho(t)} \sim - 1$, such that the expansion is exponential (See equation (\ref{w cte})). However, in $\tilde{\delta}$ Gravity the expansion rate is governed by $\mathcal{H}(t) = \frac{\dot{\tilde{R}}(t)}{\tilde{R}(t)}$. Then the accelerated expansion could be produced by a divergence in $\tilde{R}(t)$, just like we explained dark energy. Additionally in inflation, we have a new field, $\tilde{\varphi}_0$, giving us a non-zero $\tilde{\omega}(t)$. In conclusion, in inflation with $\tilde{\delta}$ Gravity, an accelerated expansion can be produced by additional factors. The application of these ideas to study inflation will be published elsewhere.\\

\newpage

\section*{Conclusions.}

We have proposed a modified model of gravity. It incorporates a new gravitational field $\tilde{g}_{\mu \nu}$ that transforms correctly under general coordinate transformations and exhibits a new symmetry: the $\tilde{\delta}$ symmetry. The new action is invariant under these transformations. We call this new gravity model $\tilde{\delta}$ Gravity. A quantum field theory analysis of $\tilde{\delta}$ Gravity has been developed in \cite{delta gravity}.\\

In this paper, we studied $\tilde{\delta}$ Gravity at a classical level. For this, we were required to set up the following three issues. First, we needed to find the equations of motion for $\tilde{\delta}$ Gravity. One of them are Einstein's equations, which  gives us $g_{\mu \nu}$, and additionally we have the equation of $\tilde{g}_{\mu \nu}$. Secondly, we needed the modified action for a test particle. This action, (\ref{Geo Action Lagr 2}), incorporates the new field $\tilde{g}_{\mu \nu}$. We obtained that a massless particle, like a photon for example, moves in a null geodesic of $\mathbf{g}_{\mu \nu} = g_{\mu \nu} +  \tilde{g}_{\mu \nu}$ and that a massive particle is governed by the equation of motion (\ref{geodesics m}). Third, we needed to fix the gauge for $g_{\mu \nu}$ and $\tilde{g}_{\mu \nu}$. For this, we developed the extended harmonic gauge given by (\ref{Harmonic gauge}) and (\ref{Harmonic gauge tilde}).\\

Then, we studied the cosmological case. In \cite{DG DE} it was shown that $\tilde{\delta}$ Gravity predicts an accelerated expansion of the Universe without a cosmological constant or additional scalar fields by using an approximation valid at small redshifts. In \cite{Paper DE}, it is developed an exact expression for the cosmological luminosity distance. However, in both cases, we assumed that we do not have $\tilde{\delta}$ matter. In this paper, we found the exact solution for $\tilde{\delta}$ Gravity with $\tilde{\delta}$ matter. For this, fixing the gauge for $g_{\mu \nu}$ and $\tilde{g}_{\mu \nu}$ was necessary, using an extended harmonic gauge. We verified that $\tilde{\delta}$ Gravity does not require dark energy to explain the accelerated expansion of the universe too. For this, we used the test particle action. We found that photons move on a null geodesic of $\mathbf{g}_{\mu \nu} = g_{\mu \nu} + \tilde{g}_{\mu \nu}$, so a new scale factor $\tilde{R}(t)$ is defined with this effective metric. In the universe, we only have non-relativistic matter and radiation, where the solution of the equations of motion is exact and $\tilde{R}(t)$ is given by (\ref{Mod Scale Factor FLRW}).\\

We also computed the age of the Universe and it is practically the same as in GR and Planck. On the other side, our model ends in a Big-Rip and we computed when it will happen. The universe has lived less than half of its life. Even though the Big-Rip could be seen as a problem, we have some way outs from the Big-Rip. Any mechanism that can provide masses to the massless particles in the model will be suitable, since then $C = 0$ which avoids a Big-Rip. For example, the appearance of massive photons at times close to the Big-Rip, by effects similar to superconductivity \cite{SuperConduct}. These effects could occur at very low temperatures which are common at later stages of the evolution of the Universe.\\

Besides, we computed the $\tilde{\delta}$ matter in the present, where we obtained that the $\tilde{\delta}$ non-relativistic matter is 23\% the ordinary non-relativistic matter. This result may imply that  dark matter is in part  $\tilde{\delta}$ matter. To verify this, it is necessary to study the CMB power spectrum with the present model. Additionally, we have a very small quantity of $\tilde{\delta}$ radiation.\\

Finally, we made a few comment about the equation of motion for only one fluid. This calculation could be useful to study more complex systems. For example, we can use them to explain the exponential expansion in inflation, just like we explained the accelerated expansion with the effective scale factor $\frac{\dot{\tilde{R}}(t)}{\tilde{R}(t)}$. Further tests of the model must include the computation of the CMB power spectrum and the formation and evolution of the large-scale structure of the universe. Work in these new directions is in progress.\\

In \cite{Alfaro 3} was noted that the Hamiltonian of $\tilde{\delta}$ models are not bounded from below, such that phantoms cosmological models \cite{phantom 1}-\cite{phantom 5}. However, it is not clear whether this problem will persist or not in a diffeomorphism-invariant model as $\tilde{\delta}$ Gravity. Phantom fields are used to explain the accelerated expansion of the Universe. However, in our model, it is produced by a small quantity of a radiation component in the Universe, not by a phantom field. Therefore, the radiation density in the present, $\Omega_R$, have to be included in the computation in spite of $\Omega_R \ll 1$. In this context, the accelerated expansion of the Universe can be interpreted as a geometric effect.\\

\newpage

\section*{Appendix A: $\tilde{\delta}$ Theories.}

In this Appendix, we will define the $\tilde{\delta}$ theories in general and their properties. For more details, see \cite{delta gravity,AppendixA}.\\

\subsection*{$\tilde{\delta}$ Variation:}

These theories consist in the application of a variation represented by $\tilde{\delta}$. As a variation, it will have all the properties of a usual variation such as:

\begin{eqnarray}
\tilde{\delta}(AB)&=&\tilde{\delta}(A)B+A\tilde{\delta}(B) \nonumber \\
\tilde{\delta}\delta A &=&\delta\tilde{\delta}A \nonumber \\
\tilde{\delta}(\Phi_{, \mu})&=&(\tilde{\delta}\Phi)_{, \mu},
\end{eqnarray}

where $\delta$ is another variation. The particular point with this variation is that, when we apply it on a field (function, tensor, etc.), it will give new elements that we define as $\tilde{\delta}$ fields, which is an entirely new independent object from the original, $\tilde{\Phi} = \tilde{\delta}(\Phi)$. We use the convention that a tilde tensor is equal to the $\tilde{\delta}$ transformation of the original tensor when all its indexes are covariant. This means that $\tilde{S}_{\mu \nu \alpha ...} \equiv \tilde{\delta}\left(S_{\mu \nu \alpha ...}\right)$ and we raise and lower indexes using the metric $g_{\mu \nu}$. Therefore:

\begin{eqnarray}
\label{tilde tensor}
\tilde{\delta}\left(S^{\mu}_{~ \nu \alpha ...}\right)
&=& \tilde{\delta}(g^{\mu \rho}S_{\rho \nu \alpha ...}) \nonumber \\
&=& \tilde{\delta}(g^{\mu \rho})S_{\rho \nu \alpha ...} + g^{\mu \rho}\tilde{\delta}\left(S_{\rho \nu \alpha ...}\right) \nonumber \\
&=& - \tilde{g}^{\mu \rho}S_{\rho \nu \alpha ...} + \tilde{S}^{\mu}_{~ \nu \alpha ...},
\end{eqnarray}

where we used that $\delta(g^{\mu \nu}) = - \delta(g_{\alpha \beta})g^{\mu \alpha}g^{\nu \beta}$.\\

\subsection*{$\tilde{\delta}$ Transformation:}

With the previous notation in mind, we can define how the tilde elements, given by (\ref{tilde tensor}), transform. In general, we can represent a transformation of a field $\Phi_i$ like:

\begin{eqnarray}
\bar{\delta} \Phi_i = \Lambda_i^j(\Phi) \epsilon_j,
\end{eqnarray}

where $\epsilon_j$ is the parameter of the transformation. Then $\tilde{\Phi}_i = \tilde{\delta}\Phi_i$ transforms:

\begin{eqnarray}
\label{tilde trans general}
\bar{\delta} \tilde{\Phi}_i = \tilde{\Lambda}_i^j(\Phi) \epsilon_j + \Lambda_i^j(\Phi) \tilde{\epsilon}_j,
\end{eqnarray}

where we used that $\tilde{\delta}\bar{\delta} \Phi_i = \bar{\delta}\tilde{\delta} \Phi_i = \bar{\delta}\tilde{\Phi}_i$ and $\tilde{\epsilon}_j = \tilde{\delta} \epsilon_j$ is the parameter of the new transformation. These extended transformations form a close algebra \cite{AppendixA}.

Now, we consider general coordinate transformations or diffeomorphism in its infinitesimal form:

\begin{eqnarray}
\label{xi 0}
x'^{\mu} &=& x^{\mu} - \xi_0^{\mu}(x) \nonumber \\
\bar{\delta} x^{\mu} &=& - \xi_0^{\mu}(x),
\end{eqnarray}

where $\bar{\delta}$ will be the general coordinate transformation from now on. Defining:

\begin{eqnarray}
\label{xi 1}
\xi_1^{\mu}(x) \equiv \tilde{\delta} \xi_0^{\mu}(x)
\end{eqnarray}

and using (\ref{tilde trans general}), we can see a few examples of how some elements transform:\\

\textbf{I)} A scalar $\phi$:

\begin{eqnarray}
\label{scalar}
\bar{\delta} \phi &=& \xi^{\mu}_0 \phi_{, \mu} \\
\label{scalar_tild}
\bar{\delta} \tilde{\phi} &=& \xi^{\mu}_1 \phi,_{\mu} + \xi^{\mu}_0 \tilde{\phi},_{\mu}.
\end{eqnarray}

\textbf{II)} A vector $V_{\mu}$:

\begin{eqnarray}
\label{vector}
\bar{\delta} V_{\mu} &=& \xi_0^{\beta} V_{\mu, \beta} + \xi_{0, \mu}^{\alpha} V_{\alpha} \\
\label{vector_tild}
\bar{\delta} \tilde{V}_{\mu} &=& \xi_1^{\beta} V_{\mu, \beta} + \xi_{1, \mu}^{\alpha} V_{\alpha} + \xi_0^{\beta} \tilde{V}_{\mu, \beta} + \xi_{0, \mu}^{\alpha} \tilde{V}_{\alpha}.
\end{eqnarray}

\textbf{III)} Rank two Covariant Tensor $M_{\mu \nu}$:

\begin{eqnarray}
\label{tensor}
\bar{\delta} M_{\mu \nu} &=& \xi^{\rho}_0 M_{\mu \nu, \rho} + \xi_{0,\nu}^{\beta} M_{\mu \beta} + \xi_{0,\mu}^{\beta} M_{\nu \beta} \\
\label{tensor_tild}
\bar{\delta} \tilde{M}_{\mu \nu}  &=& \xi^{\rho}_1 M_{\mu \nu, \rho} + \xi_{1, \nu}^{\beta} M_{\mu \beta} + \xi_{1, \mu}^{\beta} M_{\nu \beta} + \xi^{\rho}_0 \tilde{M}_{\mu \nu, \rho} + \xi_{0, \nu}^{\beta} \tilde{M}_{\mu \beta} + \xi_{0, \mu}^{\beta} \tilde{M}_{\nu \beta}.
\end{eqnarray}

These new transformations are the basis of $\tilde{\delta}$ theories. Particulary, in gravitation we have a model with two fields. The first one is just the usual gravitational field $g_{\mu \nu}$ and the second one is $\tilde{g}_{\mu \nu}$. Then, we will have two gauge transformations associated to general coordinate transformation. We will call it extended general coordinate transformation, given by:

\begin{eqnarray}
\label{trans g}
\bar{\delta} g_{\mu \nu} &=& \xi_{0 \mu ; \nu} + \xi_{0 \nu ; \mu} \\
\label{trans gt}
\bar{\delta} \tilde{g}_{\mu \nu} ( x ) &=& \xi_{1 \mu ; \nu} + \xi_{1\nu ; \mu} + \tilde{g}_{\mu \rho} \xi_{0, \nu}^{\rho} + \tilde{g}_{\nu \rho} \xi^{\rho}_{0, \mu} + \tilde{g}_{\mu \nu,\rho} \xi_0^{\rho},
\end{eqnarray}

where we used (\ref{tensor}) and (\ref{tensor_tild}). Now, we can introduce the $\tilde{\delta}$ theories.\\

\subsection*{Modified Action:}

In the last section, the extended general coordinate transformations were defined. So, we can look for an invariant action. We start by considering a model which is based on a given action $S_0[\phi_I]$ where $\phi_I$ are generic fields, then we add to it a piece which is equal to a $\tilde{\delta}$ variation with respect to the fields and we let $\tilde{\delta} \phi_J = \tilde{\phi}_J$, so that we have:

\begin{eqnarray}
\label{Action}
S [\phi, \tilde{\phi}] = S_0 [\phi] +  \int d^4x \frac{\delta S_0}{\delta \phi_I(x)}[\phi] \tilde{\phi}_I(x),
\end{eqnarray}

the index $I$ can represent any kind of indices. (\ref{Action}) give us the basic structure to define any modified element for $\tilde{\delta}$ type theories. In fact, this action is invariant under our extended general coordinate transformations developed previously. For this, see \cite{AppendixA}.\\

A first important property of this action is that the classical equations of the original fields are preserved. We can see this when (\ref{Action}) is varied with respect to $\tilde{\phi}_I$:

\begin{eqnarray}
\label{Eq_phi}
\frac{\delta S_0}{\delta \phi_I(x)}[\phi] = 0.
\end{eqnarray}

Obviously, we have new equations when varied with respect to $\phi_I$. These equations determine $\tilde{\phi}_I$ and they can be reduced to:

\begin{eqnarray}
\label{Eq_phi_tilde}
\int d^4x \frac{\delta^2 S_0}{\delta \phi_I(y) \delta \phi_J(x)}[\phi] \tilde{\phi}_J(x) = 0.
\end{eqnarray}

\newpage

\section*{Appendix B: Distances and time intervals.}

It is important to observe that the proper time is defined in terms of massive particles. The equation of motion for massive particles satisfies the important property of preserving the form of the proper time in a particle in free fall. Notice that in our case the quantity that is constant using the equation of motion for massive particles, derived from (\ref{geodesics m}), is $g_{\mu \nu}\dot{x}^{\mu}\dot{x}^{\nu}$. This single out this definition of proper time and not other. So, we must define proper time using the original metric $g_{\mu \nu}$. That is:

\begin{eqnarray}
\label{proper time}
g_{\mu \nu}\left(\frac{1}{c}\frac{dx^{\mu}}{d\tau}\right)\left(\frac{1}{c}\frac{dx^{\nu}}{d\tau}\right) = - 1 \nonumber \\
\Rightarrow d \tau = \frac{1}{c} \sqrt{-g_{\mu \nu}dx^{\mu}dx^{\nu}} \rightarrow \sqrt{-g_{0 0}} dt.
\end{eqnarray}

From here, we can see that $g_{00}<0$. On the other side, we consider the motion of light rays along infinitesimally near trajectories, using (\ref{geodesics null}) and (\ref{proper time}), to get the three-dimensional metric (See \cite{DG DE,Landau}):

\begin{eqnarray}
\label{tri metric}
d l^2 &=& \gamma_{i j}dx^{i}dx^{j} \\
\gamma_{i j} &=&  \frac{g_{0 0}}{\mathbf{g}_{0 0}}\left(\mathbf{g}_{i j} - \frac{\mathbf{g}_{i 0}\mathbf{g}_{j 0}}{\mathbf{g}_{0 0}}\right), \nonumber
\end{eqnarray}

where $\mathbf{g}_{\mu \nu} = g_{\mu \nu} + \tilde{g}_{\mu \nu}$. Therefore, we measure proper time using the metric $g_{\mu \nu}$, but the space geometry is determined by both tensor fields, $g_{\mu \nu}$ and $\tilde{g}_{\mu \nu}$. For example, in cosmology, we have:

\begin{eqnarray}
\label{tri metric cosmo}
\mathbf{g}_{\mu \nu}dx^{\mu}dx^{\nu} &=& - \left(1+3F_a(t)\right) c^2 dt^2 + R^2(t)\left(1+F_a(t)\right)\left(dx^2 + dy^2 + dz^2\right) \nonumber \\
\rightarrow d l^2 &=& R^2(t)\frac{\left(1+F_a(t)\right)}{\left(1+3F_a(t)\right)}\delta_{i j}dx^{i}dx^{j} = \tilde{R}^2(t)\delta_{i j}dx^{i}dx^{j}.
\end{eqnarray}

This means that we have the same 3-geometry as in Einstein but replacing $R(t)$ by $\tilde{R}(t)$. Therefore, in $\tilde{\delta}$ Gravity, $\tilde{R}(t)$ is the effective scale factor (it determines distances in the 3d geometry) and volume is given by $V \propto \tilde{R}^3(t)$. Besides, we have that the conditions in (\ref{tri metric cosmo}) are satisfied if $\tilde{R}^2(t) > 0$. This rule is broken when the Big-Rip is produced. Finally, using (\ref{proper time}) and (\ref{tri metric}), we can find the relation between $\tilde{R}(t)$ and redshift, $z$, given by (\ref{redshift}).\\

On the other side, using the first and second law of thermodynamics, we have that (For instance, see \cite{Padmanabhan}):

\begin{eqnarray}
dS &=& \frac{dQ}{T} = \frac{dE + pdV}{T} \nonumber \\
&=& \frac{d\left(\rho V\right) + pdV}{T} \nonumber \\
&=& \frac{V}{T}\frac{d \rho}{d T}dT + \frac{\rho+p}{T}dV,
\end{eqnarray}

where $\rho$ and $p$ are the radiation density and pressure respectively. Using $dS = 0$ for an adiabatic expansion and replacing $p = \frac{\rho}{3}$ and $\rho \propto T^4$, we obtain:

\begin{eqnarray}
\frac{dV}{V} = - 3\frac{dT}{T}.
\end{eqnarray}

Finally, we saw that $V \propto \tilde{R}^3$ in $\tilde{\delta}$ Gravity, then:

\begin{eqnarray}
T &\propto& \tilde{R}^{-1} \nonumber \\
T(z) &=& T_0(1+z),
\end{eqnarray}

where we used (\ref{redshift}).\\

\newpage

\section*{Appendix C: Perfect Fluid:}

To parametrize a perfect fluid, a usual action is \cite{PF action}:

\begin{eqnarray}
\label{Action 0 PF}
S_0 = \int d^4x \sqrt{-g} \left(\frac{R}{2\kappa} - \mathbf{r}(1 + \varepsilon(\mathbf{r})) - \lambda_1 (u^au_a +1) - \lambda_2 D_{\alpha}(\mathbf{r}U^{\alpha})\right),
\end{eqnarray}

where $\mathbf{r}$ is the number of particles per unit volume in the mean frame of reference of these particles, $\varepsilon(\mathbf{r})$ is the internal energy density per unit mass of the fluid, $u_a$ is the speed of the fluid in the local frame and $\lambda_1$ and $\lambda_2$ are Lagrange multipliers that ensure the normalization of $u_a$ and conservation of mass, respectively. Finally, we have that $U_{\alpha} = e^a_{\alpha}u_a$, where $e^a_{\alpha}$ is the Vierbein. From this action, we can see that the independent variables are $g_{\mu \nu}$, $\mathbf{r}$, $u_a$, $\lambda_1$ and $\lambda_2$, where $e^a_{\alpha}$ depend of $g_{\mu \nu}$. So, our modified action is:

\begin{eqnarray}
\label{Action PF}
S &=& \int d^4x \sqrt{-g} \left(\frac{R}{2\kappa} - \mathbf{r}\left(1 + \varepsilon(\mathbf{r})\right) - \lambda_1 \left(u^au_a + 1\right) - \lambda_2 D_{\alpha}\left(\mathbf{r}U^{\alpha}\right)\right. \nonumber \\
&& \left.- \frac{1}{2\kappa}\left(G^{\alpha \beta} - \kappa T^{\alpha \beta}\right)\tilde{g}_{\alpha \beta} + \tilde{L}_M\right) \\
\label{tilde L matter PF}
\tilde{L}_M &=& - \tilde{\mathbf{r}}\left(1 + \varepsilon(\mathbf{r}) + \mathbf{r}\varepsilon'(\mathbf{r})\right) - \tilde{\lambda}_1 \left(u^au_a + 1\right) - 2 \lambda_1 u^a\tilde{u}_a - \tilde{\lambda}_2 D_{\alpha}\left(\mathbf{r}U^{\alpha}\right) \nonumber \\
&& - \lambda_2 D_{\alpha}\left(\tilde{\mathbf{r}}U^{\alpha} + \mathbf{r}U_T^{\alpha}\right)
\end{eqnarray}

with $\tilde{\mathbf{r}} = \tilde{\delta}\mathbf{r}$, $\varepsilon'(\mathbf{r}) = \frac{\partial \varepsilon}{\partial \mathbf{r}}(\mathbf{r})$, $\tilde{u}_a = \tilde{\delta}u_a$, $U_T^{\alpha} = e^{a \alpha}\tilde{u}_a$, $\tilde{\lambda}_1 = \tilde{\delta}\lambda_1$ and $\tilde{\lambda}_2 = \tilde{\delta}\lambda_2$ are new Lagrange multipliers. The energy-momentum tensor is:

\begin{eqnarray}
\label{T PF 0}
T_{\mu \nu} = - \left(\mathbf{r}(1 + \varepsilon(\mathbf{r})) + \lambda_1 (u^au_a +1) - \lambda_{2,\alpha} \mathbf{r}U^{\alpha}\right)g_{\mu \nu} - \frac{1}{2}\lambda_{2,\alpha}\mathbf{r}\left(\delta_{\nu}^{\alpha}U_{\mu}+\delta_{\mu}^{\alpha}U_{\nu}\right)
\end{eqnarray}

and we have used $\frac{\delta e^a_{\alpha}}{\delta g_{\mu \nu}} = \frac{1}{4}\left(\delta^{\mu}_{\alpha}e^{a \nu} + \delta^{\nu}_{\alpha}e^{a \mu}\right)$. Besides, we can compute that:

\begin{eqnarray}
\label{T tilde PF 0}
\tilde{T}_{\mu \nu}
&=& - \frac{1}{4}\lambda_{2,\beta}\mathbf{r}\left(\delta_{\nu}^{\beta}U^{\alpha}\tilde{g}_{\mu \alpha} + \delta_{\mu}^{\beta}U^{\alpha}\tilde{g}_{\nu \alpha} + 2g_{\mu \nu}U^{\alpha}\tilde{g}^{\beta}_{\alpha}\right) \nonumber \\
&& - \left(\mathbf{r}(1 + \varepsilon(\mathbf{r})) + \lambda_1 (u^au_a +1) - \lambda_{2,\rho}\mathbf{r}U^{\rho}\right)\tilde{g}_{\mu \nu} \nonumber \\
&& - \frac{1}{2}\tilde{\lambda}_{2,\alpha}\mathbf{r}\left(\delta_{\nu}^{\alpha}U_{\mu}+\delta_{\mu}^{\alpha}U_{\nu}\right) - \frac{1}{2}\lambda_{2,\alpha}\tilde{\mathbf{r}}\left(\delta_{\nu}^{\alpha}U_{\mu}+\delta_{\mu}^{\alpha}U_{\nu}\right) - \frac{1}{2}\lambda_{2,\alpha}\mathbf{r}\left(\delta_{\nu}^{\alpha}U^T_{\mu}+\delta_{\mu}^{\alpha}U^T_{\nu}\right) \nonumber \\
&& - \left(\tilde{\mathbf{r}}(1 + \varepsilon(\mathbf{r}) + \mathbf{r}\varepsilon'(\mathbf{r})) + \tilde{\lambda}_1 (u^au_a +1) + 2\lambda_1 u^a\tilde{u}_a - \tilde{\lambda}_{2,\alpha} \mathbf{r}U^{\alpha} - \lambda_{2,\alpha}(\tilde{\mathbf{r}}U^{\alpha} + \mathbf{r}U_T^{\alpha})\right)g_{\mu \nu}.
\end{eqnarray}

Then, we have a modified action with ten independent variables: $g_{\mu \nu}$, $\mathbf{r}$, $u_a$, $\lambda_1$, $\lambda_2$, $\tilde{g}_{\mu \nu}$, $\tilde{\mathbf{r}}$, $\tilde{u}_a$, $\tilde{\lambda}_1$ and $\tilde{\lambda}_2$. So, we must solve (\ref{Einst Eq}) and (\ref{tilde Eq}) using (\ref{T PF 0}) and (\ref{T tilde PF 0}) to obtain $g_{\mu \nu}$ and $\tilde{g}_{\mu \nu}$. Fortunately, we can use the equations of motion for $\mathbf{r}$, $u_a$, $\lambda_1$, $\lambda_2$, $\tilde{\mathbf{r}}$, $\tilde{u}_a$, $\tilde{\lambda}_1$ and $\tilde{\lambda}_2$. These equations can be reduced to:

\begin{eqnarray}
\label{eq 1}
&& u^au_a + 1 = 0 \\
\label{eq 2}
&& D_{\alpha}(\mathbf{r}U^{\alpha}) = 0 \\
\label{eq 3}
&& 2\lambda_1 u^a - \mathbf{r}e^{a \alpha}\lambda_{2,\alpha} = 0 \\
\label{eq 4}
&& 1 + \varepsilon(\mathbf{r}) + \mathbf{r}\varepsilon'(\mathbf{r}) - U^{\alpha}\lambda_{2,\alpha} = 0 \\
\label{eq 1 tilde}
&& u^a\tilde{u}_a = 0 \\
\label{eq 2 tilde}
&& D_{\alpha}\left(\tilde{\mathbf{r}}U^{\alpha} + \mathbf{r}U_T^{\alpha} - \frac{1}{2}\mathbf{r}\tilde{g}^{\alpha \beta}U_{\beta} + \frac{1}{2}\mathbf{r}\tilde{g}_{\beta}^{\beta}U^{\alpha}\right) = 0 \\
\label{eq 3 tilde}
&& 2\tilde{\lambda}_1 u^a + 2\lambda_1 \tilde{u}^a - e^{a \alpha}\left(\mathbf{r}\tilde{\lambda}_{2,\alpha} + \tilde{\mathbf{r}}\lambda_{2,\alpha} - \frac{1}{2}\tilde{g}^{\beta}_{\alpha}\mathbf{r}\lambda_{2,\beta}\right) = 0 \\
\label{eq 4 tilde}
&& \tilde{\mathbf{r}}\left(2\varepsilon'(\mathbf{r}) + \mathbf{r}\varepsilon''(\mathbf{r})\right) - U^{\alpha}\tilde{\lambda}_{2,\alpha} - U_T^{\alpha}\lambda_{2,\alpha} + \frac{1}{2}U_{\beta}\tilde{g}^{\alpha \beta}\lambda_{2,\alpha} = 0.
\end{eqnarray}

Now, to simplify (\ref{T PF 0}) and (\ref{T tilde PF 0}), we can eliminate the Lagrange multipliers rewriting (\ref{eq 3}), (\ref{eq 4}), (\ref{eq 3 tilde}) and (\ref{eq 4 tilde}) as:

\begin{eqnarray}
\label{Id 1}
\lambda_1 &=& - \frac{1}{2}\mathbf{r}\left(1 + \varepsilon(\mathbf{r}) + \mathbf{r}\varepsilon'(\mathbf{r})\right) \\
\label{Id 2}
\lambda_{2,\mu} &=& - \left(1 + \varepsilon(\mathbf{r}) + \mathbf{r}\varepsilon'(\mathbf{r})\right)U_{\mu} \\
\label{Id 3}
\tilde{\lambda}_1 &=& - \frac{1}{2}\tilde{\mathbf{r}}\left(1 + \varepsilon(\mathbf{r}) + 3\mathbf{r}\varepsilon'(\mathbf{r}) + \mathbf{r}^2\varepsilon''(\mathbf{r})\right)\\
\label{Id 4}
\tilde{\lambda}_{2,\mu} &=&  - \tilde{\mathbf{r}}\left(2\varepsilon'(\mathbf{r}) + \mathbf{r}\varepsilon''(\mathbf{r})\right)U_{\mu} - \left(1 + \varepsilon(\mathbf{r}) + \mathbf{r}\varepsilon'(\mathbf{r})\right)\left(U^T_{\mu} + \frac{1}{2}\tilde{g}^{\beta}_{\mu}U_{\beta}\right).
\end{eqnarray}

Then, the energy-momentum tensors are:

\begin{eqnarray}
\label{T PF 1}
T_{\mu \nu} &=& \mathbf{r}^2\varepsilon'(\mathbf{r})g_{\mu \nu} + \mathbf{r}\left(1+\varepsilon(\mathbf{r})+\mathbf{r}\varepsilon'(\mathbf{r})\right)U_{\mu}U_{\nu} \\
\label{T tilde PF 1}
\tilde{T}_{\mu \nu}
&=& \mathbf{r}^2\varepsilon'(\mathbf{r})\tilde{g}_{\mu \nu} + \mathbf{r}\tilde{\mathbf{r}}\left(2\varepsilon'(\mathbf{r}) + \mathbf{r}\varepsilon''(\mathbf{r})\right)g_{\mu \nu} + \tilde{\mathbf{r}}\left(1+\varepsilon(\mathbf{r})+3\mathbf{r}\varepsilon'(\mathbf{r})+\mathbf{r}^2\varepsilon''(\mathbf{r})\right)U_{\mu}U_{\nu} \nonumber \\
&& + \mathbf{r}\left(1+\varepsilon(\mathbf{r})+\mathbf{r}\varepsilon'(\mathbf{r})\right)\left(\frac{1}{2}\left(U_{\nu}U^{\alpha}\tilde{g}_{\mu \alpha} + U_{\mu}U^{\alpha}\tilde{g}_{\nu \alpha}\right) + U^T_{\mu}U_{\nu} + U_{\mu}U^T_{\nu}\right)
\end{eqnarray}

and the equations that survive are:

\begin{eqnarray}
\label{Eq 1}
&& U^{\alpha}U_{\alpha} + 1 = 0 \\
\label{Eq 2}
&& D_{\alpha}(\mathbf{r}U^{\alpha}) = 0 \\
\label{Eq 3}
&& \left(1+\varepsilon(\mathbf{r})+\mathbf{r}\varepsilon'(\mathbf{r})\right)U^{\alpha}D_{\alpha}U_{\mu} + \left(\delta^{\alpha}_{\mu}+U^{\alpha}U_{\mu}\right)\left(2\varepsilon'(\mathbf{r})+\mathbf{r}\varepsilon''(\mathbf{r})\right)\partial_{\alpha}\mathbf{r} = 0 \\
\label{Eq 1 tilde}
&& U^{\alpha}U^T_{\alpha} = 0 \\
\label{Eq 2 tilde}
&& D_{\alpha}\left(\tilde{\mathbf{r}}U^{\alpha} + \mathbf{r}U_T^{\alpha} - \frac{1}{2}\mathbf{r}\tilde{g}^{\alpha \beta}U_{\beta} + \frac{1}{2}\mathbf{r}\tilde{g}_{\beta}^{\beta}U^{\alpha}\right) = 0 \\
\label{Eq 3 tilde}
&& \tilde{\mathbf{r}}\left(2\varepsilon'(\mathbf{r})+\mathbf{r}\varepsilon''(\mathbf{r})\right)U^{\alpha}D_{\alpha}U_{\mu} + \left(1+\varepsilon(\mathbf{r})+\mathbf{r}\varepsilon'(\mathbf{r})\right)\left(U_T^{\alpha} - \frac{1}{2}\tilde{g}^{\alpha \beta}U_{\beta}\right)D_{\alpha}U_{\mu} \nonumber \\ && + \left(1+\varepsilon(\mathbf{r})+\mathbf{r}\varepsilon'(\mathbf{r})\right)U^{\alpha}D_{\alpha}\left(U^T_{\mu} + \frac{1}{2}\tilde{g}_{\mu \beta}U^{\beta}\right) + \frac{1}{2}\left(1+\varepsilon(\mathbf{r})+\mathbf{r}\varepsilon'(\mathbf{r})\right)U^{\alpha}U^{\beta}D_{\mu}\tilde{g}_{\alpha \beta} \nonumber \\
&& + \left(\left(U_T^{\alpha} - \frac{1}{2}\tilde{g}^{\alpha \beta}U_{\beta}\right)U_{\mu} + U^{\alpha}\left(U^T_{\mu}+\frac{1}{2}\tilde{g}_{\mu \beta}U^{\beta}\right)\right)\left(2\varepsilon'(\mathbf{r})+\mathbf{r}\varepsilon''(\mathbf{r})\right)\partial_{\alpha}\mathbf{r} \nonumber \\ && + \left(\delta^{\alpha}_{\mu}+U^{\alpha}U_{\mu}\right)\left(\tilde{\mathbf{r}}\left(3\varepsilon''(\mathbf{r})+\mathbf{r}\varepsilon'''(\mathbf{r})\right)\partial_{\alpha}\mathbf{r} + \left(2\varepsilon'(\mathbf{r})+\mathbf{r}\varepsilon''(\mathbf{r})\right)\partial_{\alpha}\tilde{\mathbf{r}}\right) = 0.
\end{eqnarray}

These equations are related to (\ref{Conserv T}) and (\ref{Conserv tilde T}). So, they are a complete system of equations. Finally, from (\ref{T PF 1}) we can identify that $\rho = \mathbf{r}\left(1+\varepsilon(\mathbf{r})\right)$ and $p(\rho) = \mathbf{r}^2\varepsilon'(\mathbf{r})$. Therefore, the final expressions of the energy-momentum tensors are:

\begin{eqnarray}
\label{T PF}
T_{\mu \nu} &=& p(\rho)g_{\mu \nu} + \left(\rho + p(\rho)\right)U_{\mu}U_{\nu} \\
\label{T tilde PF}
\tilde{T}_{\mu \nu}
&=& p(\rho)\tilde{g}_{\mu \nu} + \frac{\partial p}{\partial \rho}(\rho)\tilde{\rho}g_{\mu \nu} + \left(\tilde{\rho} + \frac{\partial p}{\partial \rho}(\rho)\tilde{\rho}\right)U_{\mu}U_{\nu} \nonumber \\
&& + \left(\rho + p(\rho)\right)\left(\frac{1}{2}\left(U_{\nu}U^{\alpha}\tilde{g}_{\mu \alpha} + U_{\mu}U^{\alpha}\tilde{g}_{\nu \alpha}\right)+U^T_{\mu}U_{\nu}+U_{\mu}U^T_{\nu}\right).
\end{eqnarray}

In this paper, we will use (\ref{T PF}) and (\ref{T tilde PF}) to solve (\ref{Einst Eq}), (\ref{tilde Eq}), (\ref{Conserv T}) and (\ref{Conserv tilde T}) for a perfect fluid in the cosmological case.\\

\newpage

\section*{Appendix D: Harmonic Gauge.}

We know that the Einstein's equations do not fix all degrees of freedom of $g_{\mu \nu}$. This means that, if $g_{\mu \nu}$ is solution, then exist other solution $g'_{\mu \nu}$ given by a general coordinate transformation $x \rightarrow x'$. We can eliminate these degrees of freedom by adopting some particular coordinate system, fixing the gauge.\\

One particularly convenient gauge is given by the harmonic coordinate conditions. That is:

\begin{eqnarray}
\label{Harmonic gauge}
\Gamma^{\mu} \equiv g^{\alpha \beta}\Gamma^{\mu}_{\alpha \beta} = 0.
\end{eqnarray}

Under general coordinate transformation, $\Gamma^{\mu}$ transform:

\begin{eqnarray}
\Gamma'^{\mu} = \frac{\partial x'^{\mu}}{\partial x^{\alpha}}\Gamma^{\alpha} - g^{\alpha \beta}\frac{\partial^2 x'^{\mu}}{\partial x^{\alpha} \partial x^{\beta}}. \nonumber
\end{eqnarray}

Therefore, if $\Gamma^{\alpha}$ does not vanish, we can define a new coordinate system $x'^{\mu}$ where $\Gamma'^{\mu} = 0$. So, it is always possible to choose an harmonic coordinate system (For more detail about harmonic gauge see, for example, \cite{Weinberg grav}).\\

In the same form, we need to fix the gauge for $\tilde{g}_{\mu \nu}$. It is natural to choose a gauge given by:

\begin{eqnarray}
\label{Harmonic gauge tilde}
\tilde{\delta}\left(\Gamma^{\mu}\right) \equiv g^{\alpha \beta}\tilde{\delta}\left(\Gamma^{\mu}_{\alpha \beta}\right) - \tilde{g}^{\alpha \beta}\Gamma^{\mu}_{\alpha \beta} = 0,
\end{eqnarray}

where $\tilde{\delta}\left(\Gamma^{\mu}_{\alpha \beta}\right) = \frac{1}{2}g^{\mu \lambda}\left(D_{\beta}\tilde{g}_{\lambda \alpha}+D_{\alpha}\tilde{g}_{\beta \lambda}-D_{\lambda}\tilde{g}_{\alpha \beta}\right)$. Then, equations (\ref{Harmonic gauge}) and (\ref{Harmonic gauge tilde}) represent an extended harmonic gauge.\\


\section*{Acknowledgements.}

The work of P. Gonz\'alez has been partially financed by Beca Doctoral Conicyt $N^0$ 21080490, Fondecyt 1110378, Anillo ACT 1102, Anillo ACT 1122 and CONICYT Programa de Postdoctorado FONDECYT $N^o$ 3150398. The work of J. Alfaro is partially supported by Fondecyt 1110378, Fondecyt 1150390, Anillo ACT 1102 and Anillo ACT 11016. J.A. wants to thank F. Prada and R. Wojtak for useful remarks.\\

\newpage

\end{document}